\documentclass[preprint,showpacs,superscriptaddress,aps]{revtex4-1}
\usepackage{amssymb}
\usepackage{amsmath,amssymb,graphicx}
\usepackage{graphicx}
\usepackage{dcolumn}
\usepackage{bm}
\def\be{\begin{equation}}
\def\ee{\end{equation}}
\def\bea{\begin{eqnarray}}
\def\eea{\end{eqnarray}}

\def\lsim{\raise0.3ex\hbox{$\;<$\kern-0.75em\raise-1.1ex\hbox{$\sim\;$}}}
\def\gsim{\raise0.3ex\hbox{$\;>$\kern-0.75em\raise-1.1ex\hbox{$\sim\;$}}}

\def\bin#1#2{\left(\negthinspace\begin{array}{c}#1\\#2\end{array}\right)}

\def\fot{\frac{1}{2}}

\def\mbs{\mbox{\boldmath$\sigma$}}

\def\rb {{\bf r}}

\newcommand{\nn}{\nonumber}

\usepackage{axodraw4j}
\usepackage{pstricks}
\begin{document}

\bigskip

\vspace{2cm}
\title{Matrix elements of four-quark operators and $\Delta L=2$ hyperon decays}
\vskip 6ex
\author{C. Barbero}
\affiliation{Departamento de F\'\i sica, Universidad Nacional de la Plata, cc 67, 1900 La Plata, Argentina}
\affiliation{Instituto de F\'\i sica La Plata, CONICET, Argentina}
\author{Ling-Fong Li}
\affiliation{Department of Physics, Carnegie Mellon University, Pittsburg, Pennsylvannia 15213, USA}
\author{G. L\'opez Castro}
\email{glopez@fis.cinvestav.mx}
\affiliation{Departamento de F\'\i sica, Centro de Investigaci\'on y de Estudios Avanzados, Apartado Postal 14-740, 07000 M\'exico D.F., M\'exico}
\author{A. Mariano}
\affiliation{Departamento de F\'\i sica, Universidad Nacional de la Plata, cc 67, 1900 La Plata, Argentina}
\affiliation{Instituto de F\'\i sica La Plata, CONICET, Argentina}

\bigskip

\bigskip

\begin{center}
\begin{abstract}
  The study of neutrinoless double beta decays of nuclei and hyperons require the calculation of hadronic matrix elements of local four-quark operators that change the total charge by two units $\Delta Q=2$ . Using a low energy effective Lagrangian that induces these transitions, we compute these hadronic matrix elements in the framework of the MIT bag model. As an illustrative example we evaluate the amplitude and transition rate of $\Sigma^- \to pe^-e^-$, a decay process that violates lepton number by two units ($\Delta L=2$). The relevant matrix element is evaluated without assuming the usual factorization approximation of the four-quark operators and the results obtained in both approaches are compared.

\end{abstract}
\end{center}

\maketitle
\bigskip

\section{ Introduction}

\bigskip

 Lepton number violating (LNV) interactions with $\Delta L=2$ are widely viewed as the cleanest test of the Majorana nature of massive neutrinos \cite{Pontecorvo:1957qd}; indeed, Majorana mass terms violate lepton number by two units \cite{paul} giving rise to production or decay processes  with  $\Delta L=2$. Other mechanisms underlying the generation of neutrino masses, like the ones involving Higgs triplets \cite{cheng-li}, can also provide a source of LNV. Currently, neutrinoless double beta ($0\nu\beta\beta$) nuclear decays $(A,Z) \to (A,Z+2)e^-e^-$  have become the most sensitive probe to search for the effects of very light Majorana neutrinos \cite{doublebeta}. The underlying mechanism leading to these transitions is the conversion of two bounded neutrons in the initial nucleus into two bounded protons in the final one, making the knowledge of the nuclear wavefuntions the main limitation to achieve precise theoretical predictions. At the quark level, the elementary process $dd \to uu e^-e^-$
is responsible for $0\nu\beta\beta$ nuclear decays.

  The same simple mechanism would produce $\Delta L=2$ violation in hyperon decays, $B_1^- \to B_2^+l^-l'^-$, where $B_{1,2}$ are hyperon states and $l,l'=e$ or $\mu$. Examples of these decays are shown in Table 1:

\begin{table}[!h]
\centering
\small
\renewcommand{\arraystretch}{1.2}
\renewcommand{\arrayrulewidth}{0.8pt}
\begin{tabular}{|c|c||c|c|}
\hline
Channel & $\Delta S$ & Channel &  $\Delta S$  \\
\hline 
$\Sigma^- \to \Sigma^+e^-e^-$  & 0& $\Xi^- \to p e^-e^-$ &2\\
$\Sigma^- \to pe^-e^-$  &1& $\Xi^- \to p e^-\mu^-$ &2 \\
$\Sigma^- \to pe^-\mu^-$  &1& $\Xi^- \to p \mu^-\mu^-$  &2\\ 
$\Sigma^- \to p\mu^-\mu^-$  &1& $\Omega^- \to \Sigma^+e^-e^-$  &2\\
$\Xi^- \to \Sigma^+ e^-e^-$ &1& $\Omega^- \to \Sigma^+\mu^-e^-$ &2 \\
$\Xi^- \to \Sigma^+ \mu^-e^-$ &1& $\Omega^- \to \Sigma^+\mu^-\mu^-$ &2 \\
\hline
\end{tabular}
\caption{{\small Lepton number violating ($\Delta L=2$) decays of hyperons. The classification of these decays according to their change in strangeness ($\Delta S$) is also indicated. }}
\end{table}

Only one experimental upper limit of the channels listed in Table 1 has been reported so far, namely $B(\Xi^- \to p \mu^-\mu^-) \leq 4.0 \times 10^{-8}$ \cite{hyperCP}. A less restrictive $\Delta L=2$ decay mode in the charm sector has been reported in Ref. \cite{e653} with the following upper limit: $B(\Lambda^+_c \to \Sigma^-\mu^+\mu^+) \leq 7.4\times 10^{-4}$.  In the case of the decays listed in Table 1, two down-type ($d$ or $s$) quarks convert into two up-quarks changing the charge of hyperons according to the $\Delta Q=\Delta L=+2$ rule , as is shown in Figure 1.  These quarks conversion are assumed to occur at the same space-time location and, therefore, they are driven by local four-quark operators. Therefore, the study of the relatively simpler case provided by $0\nu\beta\beta$ hyperons decays may shed some light on the approximations used to evaluate the hadronic matrix elements relevant for similar nuclear decays. 

In the present paper we study the hadronic matrix elements of four-quark operators taken between initial and final hyperon states in the framework of the MIT bag model \cite{Chodos:1974je}. We use the effective low-energy Lagrangian proposed in Ref. \cite{Li:2007qx} which underlies $\Delta L=2$  semileptonic transitions as the ones shown in Table 1. This method provides an evaluation of the hadronic matrix elements that does not use the approximation based on the insertion of intermediate states by factorizing the four-quark operators into two quark currents. The later approximation is commonly used in the evaluation of the hadronic matrix elements in neutrinoless double-beta decays of nuclei \cite{doublebeta} and hyperons \cite{Barbero:2002wm,Barbero:2007zm}.

\bigskip

\section{Effective Lagrangian and hadronic matrix elements}

  The most general form of the low-energy effective Lagrangian that is relevant for LNV semileptonic hyperon decays was given in Ref. \cite{Li:2007qx} (the superscript $c$ labels the charge conjugated spinor):
\bea
-{\cal L}_{\beta\beta } &=& \frac{G_F^2}{\frac{}{}\Lambda_{\beta\beta}} \left\{ c_1(\bar{u}\Gamma_id)(\bar{u}\Gamma_jd)+c_2[(\bar{u}\Gamma_id)(\bar{u}\Gamma_js)+(\bar{u}\Gamma_is)(\bar{u}\Gamma_jd)]+c_3(\bar{u}\Gamma_is)(\bar{u}\Gamma_js)\right\} \nn \\ 
&& \ \ \ \ \  \times \left\{d_1(\bar{e}\Gamma_k e^c)+d_2(\bar{\mu}\Gamma_k\mu^c)+d_3(\bar{e}\Gamma_k\mu^c+\bar{\mu}\Gamma_k e^c) \right\}\ . \label{lagrangian}
\eea
Here $\Lambda_{\beta\beta}$ is a mass parameter corresponding to the physics scale for these processes and $c_i,\ d_i$ are dimensionless coefficients which represent the interaction strengths for the different channels. The dimensionless $\Gamma_i$'s are combinations of Dirac gamma matrices and depend on the physical mechanisms involved. 
The parameters and Lorentz structures involved in Eq. (\ref{lagrangian}) depend on the specific underlying model and will contain some unknown parameters \cite{Li:2007qx}. In the present paper, for the purpose of illustration, we will assume that only vector--axial structures are involved in fermionic bilinears althought it is not very difficult to consider other Lorentz structures.  
\begin{figure}[b]
  \includegraphics[width=8.5cm]{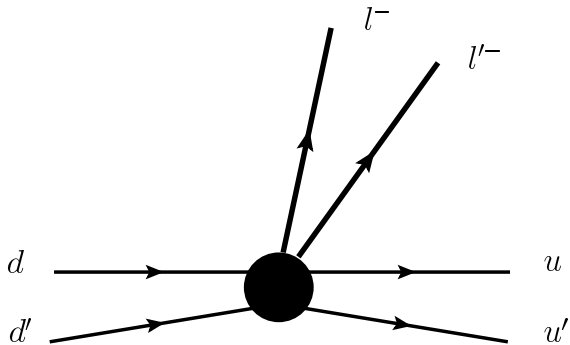}
  \caption{Feynman diagram describing the local interaction of Eq. (\ref{lagrangian}) which converts two down-type quarks into two up quarks and two leptons, $dd' \to uu'l^-l^-$.}\label{fig1}
\end{figure}

   At the lowest order in the interaction Lagrangian of Eq. (\ref{lagrangian}), the $0\nu\beta\beta$ decays of hyperons require the evaluation of the hadronic matrix elements of four-quark operators. 
The amplitude  for any of the decays listed in Table 1, which we denote as $B_1^-(p) \to B_2^+(p') l^-(p_1)l^-(p_2)$ with $l=e$ or $\mu$ (letters within brackets denote the four-momenta), is given by:
\be
{\cal M}(B_1^- \to B_2^+l^-l^-)= \frac{G_F^2}{\Lambda_{\beta\beta}}c_id_jX^{B_1 \to B_2}_{\mu\nu}L^{\mu\nu}\ ,
\ee
where 
\bea
L_{\mu\nu}&=&[\bar{u}(p_2)\gamma_{\mu}(1-\gamma_5)\gamma_{\nu}u^c(p_1)-(p_1\leftrightarrow p_2)] \nn \\
          &=& 2g_{\mu\nu}\bar{u}(p_2)(1+\gamma_5)u^c(p_1)
\eea
 is the properly antisymmetrized leptonic tensor, $c_i$ and $d_j$ are the corresponding coefficients of the operators in the Lagrangian (\ref{lagrangian})  and $u^c(p_1)$ denotes the charge conjugated of spinor $u(p_1)$; note that $L^{\mu\nu}$ becomes a symmetric tensor after using the charge conjugation property of the leptonic current \cite{Atre:2005eb}. 

\begin{figure}
  \includegraphics[width=8.5cm]{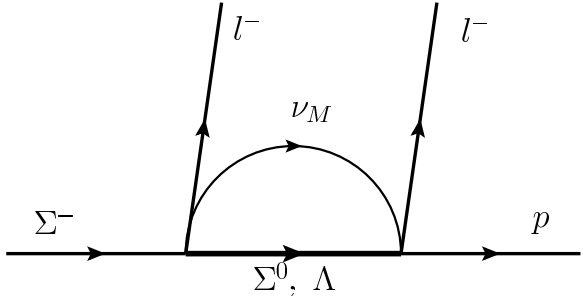}
  \caption{Feynman diagram for the $\Sigma^- \to pl^-l^-$ decay induced by the loop effect of a Majorana neutrino.}\label{fig2}
\end{figure}

The hadronic matrix element is:
\be
X^{B_1 \to B_2}_{\mu\nu}=\langle B_2^+(p')|(\bar{u}\gamma_{\mu}(1-\gamma_5)D)(\bar{u}\gamma_{\nu}(1-\gamma_5)D')|B_1^-(p) \rangle\ , \label{bme}
\ee
where $D,D'$ denote down-type quarks $d$ or $s$.
  One way to compute the hadronic matrix element is to insert, between the product of quark bilinear operators, a set of intermediate baryonic states with the appropriate quantum numbers. Usually, one has to truncate the calculation by including only a few intermediate states which are supposed to be the dominant ones (for example, the $\Sigma^0$ and $\Lambda$ hyperons in this case). This  was done in Refs.  \cite{Barbero:2002wm,Barbero:2007zm} in a model where the effects of virtual Majorana neutrinos is considered (see Figure \ref{fig2} for the specific case of $\Sigma^- \to p l^-l^-$ decay). Next, one needs to use a set of form factors to parametrize the matrix elements of weak currents at each vertex; this procedure introduces a model-dependent input in the calculations. This approximation is good as long as only a few intermediate states and the low-energy behavior of the form factors give the dominant contribution. Note however that the loop integration becomes divergent and requires the introduction of an {\it ad hoc} regulator which can be identified with some average distance between quarks inside the hyperon \cite{Barbero:2002wm,Barbero:2007zm}.

  In this paper, we use the MIT bag model of baryons to compute  the matrix element given in (\ref{bme}). Let us first note that, given the specific structure of quark currents in (\ref{bme}), we can write it as follows:
\be
\langle B_2(p') |(V-A)_{\alpha}(V'-A')_{\beta}|B_1(p) \rangle = \bar{u}(p') \left[ \Gamma_{\alpha\beta}^V-\Gamma_{\alpha\beta}^A\right]u(p)\ ,
\ee
where $u(p)$ are Dirac spinors describing the free hyperon states, and $\Gamma_{\alpha\beta}^{A,V}(P,q)$ are second-rank tensors that depends upon $P=p+p'$ and $q=p-p'$. After using Gordon identities for the vector and axial currents, the most general form of the vertices can be parametrized as:
\bea
\Gamma_{\alpha\beta}^V &=& h_1 g_{\alpha\beta} + ih_{2}\sigma_{\alpha\beta}  \nn \\ 
&& \ + \frac{h_3}{2M}\gamma_{\alpha}P_{\beta}+ \frac{h_4}{2M}\gamma_{\alpha}q_{\beta}+ \frac{h_5}{2M}\gamma_{\beta}P_{\alpha}+\frac{h_6}{2M}\gamma_{\beta}q_{\alpha} \nn \\
&& \ + \frac{h_7}{4M^2}P_{\alpha}q_{\beta} + \frac{h_8}{4M^2}q_{\alpha}q_{\beta}  + i\frac{h_{9}}{4M^2}P_{\alpha}\sigma_{\beta\mu}q^{\mu}  +i\frac{h_{10}}{4M^2}q_{\alpha}\sigma_{\beta\mu}q^{\mu} \nn \\ 
&& \  + i\frac{h_{11}}{4M^2}P_{\beta}\sigma_{\alpha\mu}q^{\mu} 
+i\frac{h_{12}}{4M^2}q_{\beta}\sigma_{\alpha \mu}q^{\mu}  + ih_{13}\epsilon_{\alpha\beta\mu\nu}\sigma^{\mu\nu}\gamma_5+
 \frac{h_{14}}{4M^2}\epsilon_{\alpha\beta\mu\nu}P^{\mu} q^{\nu}\gamma_5\ \nn \\
&& \ + \frac{h_{15}}{2M} \epsilon_{\alpha\beta\mu\nu}q^{\mu}\gamma^{\nu}\gamma_5 + 
\frac{h_{16}}{2M} \epsilon_{\alpha\beta\mu\nu}P^{\mu}\gamma^{\nu}\gamma_5\ ,
\eea 
where $M=(m+m')/2$, with $m(m')$ the  mass of the initial(final) hyperon state. The coefficients $h_i$ are $q^2$-dependent form factors which depend on the specific $B_1 \to B_2$ transition.  
Similarly, the axial vertex can be obtained by means of the following replacement: $\Gamma_{\alpha\beta}^A=\Gamma_{\alpha\beta}^V(h_i \rightarrow g_i)\times \gamma_{5}$. The form factors $h_i, g_i$ in the vector and axial vertices have all a common dimension of energy. The contributions proportional to $q/M$ are suppressed and the terms containing the Levi-Civita tensors do not contribute to the decay amplitude because of the symmetric leptonic tensor in Eq. (2).

\bigskip

\section{Form factors in the MIT bag model}

 For definiteness, let us consider the specific example of the $\Sigma^- \to pe^-e^-$ transition; in this case only the operators with coefficients $c_2$ and $d_1$ in Eq. (\ref{lagrangian}) give a contribution. In the framework of the MIT bag model, the vector and axial components of the hadronic matrix element in the $\Sigma^- \to p$ transition, Eq. (4), can be written as follows:
\bea
X_{\alpha\beta}^{\Sigma^- \to p}(V)&=& \bar{u}(p')\Gamma_{\alpha\beta}^Vu(p) \nn \\
&=& \langle p|
[M_{\alpha\beta}^{ds\rightarrow uu}+M_{\alpha\beta}^{sd\rightarrow uu}]
+[Q_{\alpha\beta}^{ds\rightarrow uu}+Q_{\alpha\beta}^{sd\rightarrow uu}]
|\Sigma^-\rangle  \ ,\\
X_{\alpha\beta}^{\Sigma^- \to p}(A)&=&  \bar{u}(p')\Gamma_{\alpha\beta}^Au(p) \nn \\
&=& \langle p|
[N_{\alpha\beta}^{ds\rightarrow uu}+N_{\alpha\beta}^{sd\rightarrow uu}]
+[P_{\alpha\beta}^{ds\rightarrow uu}+P_{\alpha\beta}^{sd\rightarrow uu}]
|\Sigma^-\rangle \ , 
\eea
where we have defined (latin indices $a,b,c,d$ denote flavor labels):
\bea
M_{\alpha\beta}^{bd\rightarrow ac}&=&\int d^3 x [\bar{\psi}_a(x)\gamma_\alpha \psi_b(x)]\cdot [\bar{\psi}_c(x)\gamma_\beta \psi_d(x)],\nn\\ 
N_{\alpha\beta}^{bd\rightarrow ac}&=&\int d^3 x [\bar{\psi}_a(x)\gamma_\alpha \psi_b(x)]\cdot [\bar{\psi}_c(x)\gamma_\beta \gamma_5\psi_d(x)],\nn\\
P_{\alpha\beta}^{bd\rightarrow ac}&=&\int d^3 x [\bar{\psi}_a(x)\gamma_\alpha\gamma_5 \psi_b(x)]\cdot [\bar{\psi}_c(x)\gamma_\beta \psi_d(x)],\nn\\
Q_{\alpha\beta}^{bd\rightarrow ac}&=&\int d^3 x [\bar{\psi}_a(x)\gamma_\alpha\gamma_5 \psi_b(x)]\cdot [\bar{\psi}_c(x)\gamma_\beta \gamma_5 \psi_d(x)]\ .
\label{20}\eea
In the above expresions $\psi_i(x)$ denotes the wavefunction of quark with flavor $i$ in the MIT bag model which is calculated according to Ref. \cite{Chodos:1974je} and it is reproduced in the Appendix.

  In the non-relativistic limit for baryons, the only non-vanishing matrix elements turn out to be the following:
\bea
X_{00}^{\Sigma^- \to p}(V)
&=&\langle p|[M_{00}^{ds\rightarrow uu}+M_{00}^{sd\rightarrow uu}]|\Sigma^-\rangle \nn \\
&=& {\cal R}(w_{1-1},w_{1-1},R)\left[\langle p|\sum_j\beta^+_j\sum_i\tau^+_i|\Sigma^-\rangle +\langle p|\sum_i\tau^+_i\sum_j\beta^+_j|\Sigma^-\rangle \right]. \\
X_{0k}^{\Sigma^- \to p}(A)\!\! &=&\!\!-\langle p|[N_{0k}^{ds\rightarrow uu}+N_{0k}^{sd\rightarrow uu}]|\Sigma^-\rangle \nn \\
\!\! &=&\!\! -{\cal S}(w_{1-1},w_{1-1},R)\left[\langle p|\sum_j\beta^+_j\sum_i\sigma_{k,i}\tau^+_i|\Sigma^-\rangle+
\langle p|\sum_j\tau^+_j\sum_i\sigma_{k,i}\beta^+_i|\Sigma^-\rangle\right] \\
X_{k0}^{\Sigma^- \to p}(A)&=&-{\cal S}(w_{1-1},w_{1-1},R)\left[\langle p|\sum_j\sigma_{k,j}\beta^+_j\sum_i\tau^+_i|\Sigma^-\rangle+\langle p|\sum_j\sigma_{k,j}\tau^+_j\sum_i\beta^+_i|\Sigma^-\rangle\right]. \\
X_{jk}^{\Sigma^- \to p}(V)&=&\left\{{\cal T}(w_{1-1},w_{1-1},R)\left[\langle p|\sum_l\sigma_{j,l}\beta^+_l\sum_i\sigma_{k,i}\tau^+_i|\Sigma^-\rangle+
\langle p|\sum_l\sigma_{j,l}\tau^+_l\sum_i\sigma_{k,i}\beta^+_i|\Sigma^-\rangle \right]\right.\nn\\
&&\ \ +\delta_{jk}\left.{\cal U}(w_{1-1},w_{1-1},R)\left[\langle p|\sum_l\beta^+_l\sum_i\tau^+_i|\Sigma^-\rangle +
\langle p|\sum_l\tau^+_l\sum_i\beta^+_i|\Sigma^-\rangle \right]\right\}\ . 
\label{57}
\eea
In the previous expressions $\tau^{\pm}_i$, $\beta^{\pm}_i$ denote, respectively, the isospin and $U$-spin
raising/lowering operators acting over the quark states in position $i$ within the spin-flavor wavefunctions of $|\Sigma^-\rangle $ and $|p\rangle$ (see Ref. \cite{kokkedee}). Similarly, $\sigma_{k,j}$ refer to the $k$-th component of the spin operator acting on the quark state in position $j$ in the hyperon spin-flavor wavefunction. 
On the other hand, the functions ${\cal R}, {\cal S}, {\cal T}$ and ${\cal U}$ introduced in Eqs. (10)-(13) arise from the integration over spatial coordinates of the quark wavefunctions in the MIT bag model; they depend upon the bag model parameters as shown in the expressions given in the Appendix.

  The matrix elements in Eq. (10)-(\ref{57}) can be readily evaluated by using the quark model spin-flavor wavefunctions \cite{kokkedee} of the $\Sigma^-$ and $p$ states.  An explicit calculation yields:
\bea
\langle p| \sum_j\beta_j^+\sum_i\tau_i^+|\Sigma^- \rangle &=& \langle p| \sum_i\tau_i^+ \sum_j\beta_j^+|\Sigma^- \rangle =1 \nn \\
\langle p |\sum_j\beta_j^+\sum_i\sigma_{k,i}\tau_i^+|\Sigma^-\rangle &=& \langle p |\sum_i\sigma_{k,i}\tau_i^+\sum_j\beta_j^+|\Sigma^-\rangle= \frac{5}{3}\delta_{k3} \nn \\
\langle p |\sum_j\sigma_{k,j}\beta_j^+\sum_i\tau_i^+ |\Sigma^-\rangle &=& \langle p |\sum_i\tau_i^+\sum_j\sigma_{k,j}\beta_j^+ |\Sigma^-\rangle= - \frac{1}{3}\delta_{k3} \ ,\nn
\eea
and it follows (hereafter we omit the superscript labels of the hadronic matrix elements)
\bea
X_{00}(V)&=&2{\cal R}(w_{1-1},w_{1-1},R). \\
X_{0k}(A)&=&-\frac{4}{3}\delta_{k,3}{\cal S}(w_{1-1},w_{1-1},R) \\
X_{k0}(A)&=&-\frac{4}{3}\delta_{k,3}{\cal S}(w_{1-1},w_{1-1},R) \\
X_{jk}(V)
&=&2\left\{\frac{1}{3}{\cal T}(w_{1-1},w_{1-1},R)\left[\delta_{j,3}\delta_{k,3}-(1-\delta_{j,3})(1-\delta_{k,3})i^{j+k}
[2(-1)^{j+1}+(-1)^k]\right]\right.\nn\\
&&\ \ + \frac{}{}\delta_{jk}\left.{\cal U}(w_{1-1},w_{1-1},R)\right\}\ .
\label{58}
\eea 

  On the other hand, taking the non-relativistic limit of Eqs. (5) and (6) we get the following expressions for the non-vanishing hadronic matrix elements:
\bea
X_{00}(V)&=& h_1+h_3+h_5 \ \equiv \ f_1 \nn \\
X_{0k}(A)&=&- (g_2-g_5+2ih_{13})\delta_{k,3}\ \equiv \ f_2\delta_{k,3},\nn\\
X_{k0}(A)&=& (g_2+g_3+2ih_{13})\delta_{k,3}\ \equiv \  f_3\delta_{k,3},\nn\\
X_{jk}(V)&=& -\delta_{jk}h_1+(ih_2-h_{16})(\delta_{j,1}\delta_{k,2}-\delta_{j,2}\delta_{k,1})
\ \equiv \  f_4\delta_{jk}+f_5(\delta_{j,1}\delta_{k,2}-\delta_{j,2}\delta_{k,1})\ ,
\label{gab4}
\eea
or equivalently, from Eqs. (17) and (18):
\bea
f_1&=&2{\cal R}(w_{1-1},w_{1-1},R)\ , \nn \\
f_2&=&f_3=-\frac{4}{3}{\cal S}(w_{1-1},w_{1-1},R)\ , \nn \\
f_4&=&\frac{2}{3}{\cal T}(w_{1-1},w_{1-1},R)+2{\cal U}(w_{1-1},w_{1-1},R)\ , \nn \\
f_5&=&2i{\cal T}(w_{1-1},w_{1-1},R)\ ,
\label{69}
\eea
for the effective form factors $f_i$. Note that, in the non-relativistic limit, all these form factors should be evaluated at zero momentum transfer ($q^2=0$). Thus, we are not able to provide their momentum dependence; however, as in the case of the beta decays of hyperons, we may expect that these $q$-dependent effects would affect the decays rates by at most 10$\sim$20\% given that they are a SU(3) symmetry breaking scale of order $q/M$ \cite{augusto}. 

  The contraction of Lorentz indices in Eqs. (2)--(3) leads to the following simple form of the decay amplitude:
\be
{\cal M}(\Sigma^- \to pe^-e^-)= \frac{G_F^2}{\Lambda_{\beta\beta}}c_2d_1 2\bar{u}(p_2)(1+\gamma_5)u^c(p_1)\cdot \bar{u}(p')\left[A+B\gamma_5 \right]u(p)\ ,
\ee
where $A=4h_1+h_3+h_5=f_1-3f_4$ and $B=4g_1+g_4+g_6$. In the non-relativistic limit described above the numerical evaluations of Eqs. (\ref{69}) obtained in the framework of the MIT bag model lead to $A= 3.56\times 10^5$ MeV$^3$  and $B=0$ (note the simplified expressions given in the Appendix for the functions ${\cal R},\ {\cal S},\ {\cal T}$ and ${\cal U}$). We have used the numerical values of the bag model parameters given in Ref. \cite{Chodos:1974je}.

From the above decay amplitude we get the following expression for the differential decay rate:
\bea
\frac{d\Gamma}{dq^2}&=& \frac{G_F^4}{32\pi^3m_{\Sigma}^3} \left(\frac{c_2d_1}{\Lambda_{\beta\beta}} \right)^2 
(q^2-2m_l^2) (u^{max}-u^{min}) \nonumber \\
&& \ \ \ \ \ \ \times \left\{ A^2[(m_{\Sigma}+m_p)^2-q^2]+B^2[(m_{\Sigma}-m_p)^2-q^2] \right\}\ ,
\eea
where $u^{max}-u^{min}=(\sqrt{1-4m_e^2/q^2})\cdot \lambda^{1/2}(m_{\Sigma}^2, m_p^2, q^2)$ and $\lambda(x,y,x)$ denotes the triangle function. By using $\tau_{\Sigma^-}=1.479\times 10^{-10}$s \cite{Beringer:1900zz}, and integrating numerically the differential rate in the range $4m_e^2 \leq q^2 \leq (m_{\Sigma}-m_p)^2$, we get the branching ratio for the di-electron channel:
\be
B^{\rm bag}(\Sigma^- \to pe^-e^-)=\left(\frac{c_2d_1}{\Lambda_{\beta\beta}} \right)^2\cdot (4.65\times 10^{-13}\ {\rm MeV}^2)  \ . \label{bagbr}
\ee
If the take for weak coupling coefficients $c_i,d_i \sim O(1)$ and $\Lambda_{\beta\beta}\geq 100$ GeV, the branching fraction turns out to be extremely suppressed: $B^{\rm bag}(\Sigma^- \to pe^-e^-) \leq 10^{-23}$. 

Equivalently, we can define the following ratio \cite{Li:2007qx}:
\be
R_{ee/e\nu}(\Sigma^-)\equiv \frac{\Gamma(\Sigma^- \to pe^-e^-)}{\Gamma(\Sigma^- \to ne^-\bar{\nu}_e)} .\label{ratio}
\ee
Using the result given in (\ref{bagbr}), we get $R_{ee/e\nu}^{\rm bag}(\Sigma^-) \approx (c_2d_1/\Lambda_{\beta\beta})^2 \cdot (4.6\times 10^{-10}\ {\rm MeV}^2)$. If we compare this number with the estimate given in Ref. \cite{Li:2007qx}, which is based on pure dimensional arguments, the result of our present calculation appears to be smaller by six orders of magnitude.  On the other hand, the corresponding result obtained by using the approximation based on the insertion of intermediate baryon states in the baryon-neutrino loop of Figure 2 is $R_{ee/e\nu}^{\rm loop}(\Sigma^-)=7.2 \times 10^{-18} |\langle m_{ee} \rangle|^2$ MeV$^{-2}$ (from Table 1 in Ref. \cite{Barbero:2007zm}).  Using the current limit $|\langle m_{ee} \rangle| \leq 1$ eV, we get $R_{ee/e\nu}^{loop}(\Sigma^-)\leq 7.2 \times 10^{-30}$ or, equivalently, a branching fraction $B^{\rm loop}(\Sigma^- \to p e^-e^-)\leq 7.3 \times 10^{-33}$. This upper limit on $R_{ee/e\nu}^{loop}(\Sigma^-)$ is smaller than the one obtained in our bag model calculation by ten orders of magnitude (assuming $c_2=d_1\sim O(1)$ and $\Lambda_{\beta\beta} \geq 100$ GeV). Note however, that the result of Ref. \cite{Barbero:2007zm} depends strongly on the cutoff used to regularize the integral over the virtual neutrino momentum in the loop.

 \bigskip 

\section{Conclusions}

  In this paper we have used the MIT bag model \cite{Chodos:1974je} to evaluate the hadronic matrix elements of four-quark operators required to compute the rates of $0\nu\beta\beta$ decays of hyperons. These four-quark operators appears in the most general $\Delta Q=2$ low-energy effective Lagrangian \cite{Li:2007qx} that describes the $\Delta L=2$ lepton number violation in hyperon transitions. This method avoids the use of the approximation based on the insertion of baryon intermediate states, which requires the knowledge of form factors for hyperon beta decays at very high energy scales. To our knowledge, this is a novel method for  direct calculations of hadronic matrix elements of four-quark operators. 

  As an specific example, we have considered the $\Sigma^- \to pe^-e^-$ lepton number violating decay and have computed the non-vanishing form factors of the $\Sigma^- \to p$ transition in the non-relativistic limit using the spin-flavor wavefunctions of the hyperon states. Using reasonable values for the order of magnitude of couplings and mass scales of the underlying New Physics, we have computed the upper limit on the branching ratio which turns out to be of order $10^{-23}$ for the $\Sigma^- \to pe^-e^-$ decay. 
Althought this result turns out to be ten orders of magnitude larger that the calculation based on a model where this decay is induced by a loop of baryons and light Majorana neutrinos \cite{Barbero:2007zm}, it is still very small to be accessible to sensitivities reached by current experiments. It shows, however, that the calculations based on models involving loops of virtual neutrinos and the insertion of virtual intermediate hyperon states may underestimate the true branching fractions.

\bigskip

\subsection*{Acknowledgements}

GLC acknowledges financial support from Conacyt (M\'exico). C.B. and A.M. are fellows of CONICET,
CCT La Plata (Argentina), and have received support under Grant PIP No. 349.

\bigskip

\appendix{\bf Appendix}

  In the MIT bag model the eigenfunctions of quarks confined within a baryon, which is assumed to be a spherical bag of radius $R$, are given by \cite{Chodos:1974je} (note that $\kappa=\pm 1$):
\be
\psi_{n\kappa m}(x,t)=\frac{1}{\sqrt{4\pi}}\bin{ij_{\fot(\kappa+1)}(w_{n\kappa}r/R) (\mbs\cdot \hat{\rb})^{\fot(\kappa+1)}  U_m}
{(-1)^{\fot(1-\kappa)}j_{\fot(1-\kappa)}(w_{n\kappa}r/R) (\mbs\cdot \hat{\rb})^{\fot(1-\kappa)} U_m}
e^{-iw_{n\kappa}t/R}\ ,
\label{A1}
\ee
where $w_{n\kappa}$ satisfies the eigenvalue condition $\tan w_{n\kappa}=w_{n\kappa}/(1+w_{n\kappa})$, $j_{\fot(\kappa\pm 1)}$ are spherical Bessel functions and $U_m$ are two-component Pauli spinors. 
These eigenfunctions are normalized according to  
\be
\int d^3x N(w_{n\kappa})N(w_{n'\kappa'}) \psi^\dagger_{n\kappa
m}(x,t)\psi_{n'\kappa' m}(x,t)=\delta_{nn'}\delta_{\kappa\kappa'}\delta_{mm'}\ ,
\label{A10}
\ee
and the normalization factors are defined from the following integrals:
\bea
\int_0^R r^2 dr j^2_0(w_{n\kappa}r/R)&=&
\frac{1}{4N^2(w_{n\kappa})}\frac{2w_{n\kappa}+\kappa}{w_{n\kappa}+\kappa},
\nn\\ 
\int_0^R r^2 dr j^2_1(w_{n\kappa}r/R)&=&
\frac{1}{4N^2(w_{n\kappa})}\frac{2w_{n\kappa}+3\kappa}{w_{n\kappa}+\kappa}\ .
\label{A8}
\eea

  The general form of the integrals involving the  product of four eigenfunctions required in our calculations of hadronic matrix elements, see Eq. (9), are 
\bea
\int d^3x \psi^\dagger_{L_1}(x,t)\psi_{L_2}(x,t)
\psi^\dagger_{L_3}(x,t)\psi_{L_4}(x,t)&=&\frac{\widehat{\delta}_{n\kappa }}{N_{1234}}\delta_{m_1m_2}\delta_{m_3m_4}{\cal R}(w_{n_1\kappa_1},w_{n_3\kappa_3},R) , \\ 
\int d^3x 
\psi^\dagger_{L_1}(x,t)\psi_{L_2}(x,t)
\psi^\dagger_{L_3}(x,t)\sigma_k\psi_{L_4}(x,t) 
&=&\frac{\widehat{\delta}_{n\kappa }}{N_{1234}}\delta_{m_1m_2}\sigma_k^{34} {\cal S}(w_{n_1\kappa_1},w_{n_3\kappa_3},R), \\ 
\int d^3x 
\psi^\dagger_{L_1}(x,t)\sigma_j\psi_{L_2}(x,t)
\psi^\dagger_{L_3}(x,t)\sigma_k\psi_{L_4}(x,t)
 &=&\frac{\widehat{\delta}_{n\kappa }}{N_{1234}}\times \left[
\sigma_j^{12}\sigma_k^{34} {\cal T}(w_{n_1\kappa_1},w_{n_3\kappa_3},R)\right. \nn \\  
&&\left. \ \  +\delta_{m_1m_2}\delta_{m_3m_4}{\cal U}(w_{n_1\kappa_1},w_{n_3\kappa_3},R)\right]
\label{C11}
\eea
where we have introduced the following notation: $N_{1234}\equiv N(w_{n_1\kappa_1})N(w_{n_2\kappa_2})N(w_{n_3\kappa_3})N(w_{n_4\kappa_4})$, $L_i=n_i\kappa_i m_i$, $\sigma_k^{ij}\equiv U_{m_i}^\dagger \sigma_k U_{m_j}$ and $\widehat{\delta}_{n\kappa }\equiv \delta_{n_1n_2}\delta_{n_3n_4}\delta_{\kappa_1\kappa_2}\delta_{\kappa_3\kappa_4}$. 

  By inserting the solutions given in Eq. (\ref{A1}) in the previous results, one gets:
\bea
\frac{4\pi}{N_{1133}}{\cal R}(w_{n_1\kappa_1},w_{n_3\kappa_3},R)&=& \int_0^Rr^2dr\left[j_0^2\left(\frac{w_{n_1\kappa_1}r}{R}\right)+j_1^2\left(\frac{w_{n_1\kappa_1}r}{R}\right) \right] \nn \\
&& \ \ \ \  \times \left[j_0^2\left(\frac{w_{n_3\kappa_3}r}{R}\right)+j_1^2\left(\frac{w_{n_3\kappa_3}r}{R}\right) \right] \\
\frac{4\pi}{N_{1133}}{\cal S}(w_{n_1\kappa_1},w_{n_3\kappa_3},R)&=& \int_0^Rr^2dr\left[j_0^2\left(\frac{w_{n_1\kappa_1}r}{R}\right)+j_1^2\left(\frac{w_{n_1\kappa_1}r}{R}\right) \right] \nn \\
&& \ \ \ \  \times \left[j_0^2\left(\frac{w_{n_3\kappa_3}r}{R}\right)-\frac{1}{3}j_1^2\left(\frac{w_{n_3\kappa_3}r}{R}\right) \right] \\
\frac{4\pi}{N_{1133}}{\cal T}(w_{n_1\kappa_1},w_{n_3\kappa_3},R)&=& \int_0^Rr^2dr\left[j_0^2\left(\frac{w_{n_1\kappa_1}r}{R}\right)j_0^2\left(\frac{w_{n_3\kappa_3}r}{R}\right)-\frac{1}{3}j_0^2\left(\frac{w_{n_1\kappa_1}r}{R}\right)j_1^2\left(\frac{w_{n_3\kappa_3}r}{R}\right) \right. \nn \\
&& \left. -\frac{1}{3}j_1^2\left(\frac{w_{n_1\kappa_1}r}{R}\right)j_0^2\left(\frac{w_{n_3\kappa_3}r}{R}\right)-\frac{1}{3}j_1^2\left(\frac{w_{n_1\kappa_1}r}{R}\right)j_1^2\left(\frac{w_{n_3\kappa_3}r}{R}\right) \right] \\
\frac{4\pi}{N_{1133}}{\cal U}(w_{n_1\kappa_1},w_{n_3\kappa_3},R)&=&
\int_0^Rr^2dr\left[\frac{4}{3} j_1^2\left(\frac{w_{n_1\kappa_1}r}{R}\right)j_1^2\left(\frac{w_{n_3\kappa_3}r}{R}\right) \right]\ .
\eea
A numerical evaluation of these radial integrals leads to the values of the the form factors in Eqs. (19).

\end{document}